\documentclass{raa}

\usepackage[varg]{txfonts}

\usepackage{natbib}
\usepackage{graphicx}
\usepackage{epstopdf}
\usepackage{amssymb}
\usepackage{float}
\usepackage{longtable}

\bibpunct{(}{)}{;}{a}{}{,}

\begin{document}

\title{Photometric investigation on the W-subtype contact binary V1197 Her}

   \volnopage{Vol.0 (200x) No.0, 000--000}      %%preserved for Editor. DOn't remove!
   \setcounter{page}{1}          %%starting page, preserved for Editor. DOn't remove!

\author{Zhou X.
      \inst{1,2,3,4}
   \and Soonthornthum B.
      \inst{2}
}
%%%

\institute{
             Yunnan Observatories, Chinese Academy of Sciences (CAS), P.O. Box 110, 650216 Kunming, P. R. China {\it zhouxiaophy@ynao.ac.cn}\\
%% Please give the E-mail address of the author, to whom future correspondence and
%% offprint requests will be sent.
        \and
             National Astronomical Research Institute of Thailand, 260  Moo 4, T. Donkaew,  A. Maerim, Chiangmai, 50180, Thailand\\
        \and
             Key Laboratory of the Structure and Evolution of Celestial Objects, Chinese Academy of Sciences, P. O. Box 110, 650216 Kunming, P. R. China\\
        \and
             Center for Astronomical Mega-Science, Chinese Academy of Sciences, 20A Datun Road, Chaoyang Dis-trict, Beijing, 100012, P. R. China\\
}

\date{Received~~2017 month day; accepted~~2017~~month day}

\def\gsim{\mathrel{\raise.5ex\hbox{$>$}\mkern-14mu
                \lower0.6ex\hbox{$\sim$}}}

\def\lsim{\mathrel{\raise.3ex\hbox{$<$}\mkern-14mu
               \lower0.6ex\hbox{$\sim$}}}

\abstract{
Multi-color light curves of V1197 Her were obtained with the 2.4 meter optical telescope at Thai National Observatory and the Wilson-Devinney (W-D) program is used to model the observational light curves. The photometric solutions reveal that V1197 Her is a W-subtype shallow contact binary system with a mass ratio of $q = 2.61 $ and fill-out factor to be $f = 15.7\,\%$. The temperature difference between the primary star and secondary star is only $140K$ in spite of the low degree of contact, which means that V1197 Her is not only in geometrical contact configuration but also already under thermal contact status. The orbital inclination of V1197 Her is as high as $i = 82.7^{\circ}$, and the primary star is completely eclipsed at the primary minimum. The totally eclipsing characteristic implies that the determined physical parameters are highly reliable. The masses, radii and luminosities of the primary star (star 1) and secondary star (star 2) are estimated to be $M_{1} = 0.30(1)M_\odot$, $M_{2} = 0.77(2)M_\odot$, $R_{1} = 0.54(1)R_\odot$, $R_{2} = 0.83(1)R_\odot$, $L_{1} = 0.18(1)L_\odot$ and $L_{2} = 0.38(1)L_\odot$. The evolutionary status of the two component stars are drawn in the H - R diagram, which shows that the less massive but hotter primary star is more evolved than the secondary star. The period of V1197 Her is decreasing continuously at a rate of $dP/dt=-2.58\times{10^{-7}}day\cdot year^{-1}$, which can be explained by mass transfer from the more massive star to the less massive one with a rate of $\frac{dM_{2}}{dt}=- 1.61\times{10^{-7}}M_\odot/year$. The light curves of V1197 Her is reported to have the O'Connell effect. Thus, a cool spot is added to the massive star to model the asymmetry on light curves.
}

\authorrunning{Zhou et al.}
\titlerunning{W-subtype contact binary}

\maketitle

%% From the front matter, we move on to the body of the paper.
%% In the first two sections, notice the use of the natbib \citep
%% and \citet commands to identify citations.  The citations are
%% tied to the reference list via symbolic KEYs. The KEY corresponds
%% to the KEY in the %\bibitem in the reference list below. We have
%% chosen the first three characters of the first author's name plus
%% the last two numeral of the year of publication as our KEY for
%% each reference.

\section{Introduction}\label{intro}

Contact binaries, also known as W UMa type binaries, are typical interaction binaries in the Universe. The evolutionary scenario of contact binaries are quite different from singe stars due to the existence of common envelope surrounding component stars in contact binary systems. Over the past few decades, the photometric and spectrometric data on contact binaries are accumulating rapidly owe to the large sky survey projects such as the Optical Gravitational Lensing Experiment (OGLE) \citep{1992AcA....42..253U}, the All Sky Automated Survey (ASAS) \citep{1997AcA....47..467P}, the Wide Angle Search for Planets (SuperWASP) \citep{2003ASPC..294..405S}, the Large Sky Area Multi-Object Fibre Spectroscopic Telescope (LAMOST) \citep{2012RAA....12..723Z,2017RAA....17...87Q,2019RAA....19...64Q} and the Gaia mission \citep{2016A&A...595A...1G,2018A&A...616A...1G}. Contact binaries are useful tools in study a wide range of astrophysical problems (eg. testing stellar evolution models, searching for peculiar stars) and is becoming more and more important.

V1197 Her is an EW type eclipsing binary with a period of $P = 0.262680$ days listed in the International Variable Star Index (VSX) \citep{2006SASS...25...47W}. The target has been neglected since it was discovered. Neither radial velocity nor light curve was presented before. Only surface effective temperature is derived by the Gaia mission, published by the Gaia Data Release 2 as $T = 4973K$. In the present work, we are going to study the contact binary V1197 Her thoroughly for the first time. Observations of complete light curves and mid-eclipse times are illustrated in Section 2. Analyses on its period variations are in Section 3. Modeling of light curves are in Section 4. Discussion and conclusion are made in the last section.

\section{OBSERVATIONS AND DATA REDUCTION}

\begin{figure}[!ht]
\begin{center}
\includegraphics[width=12cm]{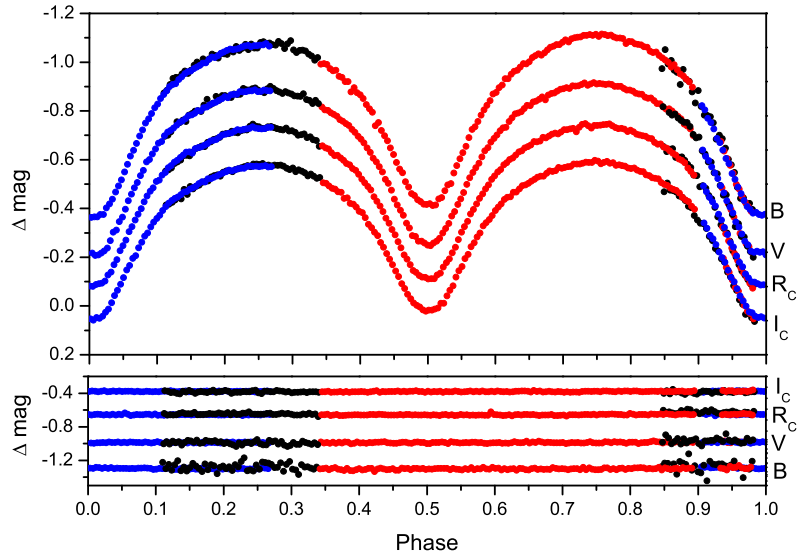}
\caption{The black, red and blue colors refer to data obtained on 20 April 2016, 25 April 2016 and 9 April 2019, respectively. The $BVR_CI_C$ band light curves of V1197 Her are shown in the up panel and magnitude differences between the Comparison star and Check star are displayed in the bottom panel. }\label{LC_Obs}
\end{center}
\end{figure}

The complete light curves of V1197 Her were obtained on 20 April 2016 , 25 April 2016 and 9 April 2019 with the 2.4 meter optical telescope at Thai National Observatory (TNO 2.4m), National Astronomical Research Institute of Thailand (NARIT). The field of view (FOV) is 16 $\times$ 16 square arc-minutes and a 4K $\times$ 4K CCD camera is equipped in the terminal \citep{2018NatAs...2..355S}. Johnson-Cousins' $BVR_CI_C$ filters were used during the observations. All observational images were reduced with the Image Reduction and Analysis Facility (IRAF) \citep{1986SPIE..627..733T}. UCAC4 592-055825 and UCAC4 592-055831 in the same FOV were selected as the Comparison (C) star and Check star (Ch) as differential photometry method was adopted to determine the light variations of V1197 Her. Their coordinates in J2000.0 epoch and $V$ band magnitudes are listed in Table \ref{Coordinates}. The observational light curves are displayed in Fig. \ref{LC_Obs}. The Heliocentric Julian Date (HJD) are converted to phases with the following equation:

\begin{equation}
Min.I(HJD) = 2458583.2589+0^{d}.262680\times{E}.\label{phase}
\end{equation}

The initial epoch used in Equation \ref{phase} was obtained in 9 April 2019 and parabola fit on the observed light minimum was performed. In all, we got two mid-eclipse times with the TNO 2.4m, which are listed in Table \ref{New_minimum}.

\begin{table}[!h]\small
\begin{center}
\caption{Coordinates and $V$ band magnitudes.}\label{Coordinates}
\begin{tabular}{cccc}\hline\hline
Target              &   $\alpha_{2000}$         &  $\delta_{2000}$          &  $V_{mag}$   \\ \hline\hline
V1197 Her           &$16^{h}33^{m}22^{s}.855$   & $+28^\circ18'19''.475$    &  $13.378$     \\
UCAC4 592-055825    &$16^{h}33^{m}20^{s}.995$   & $+28^\circ16'16''.089$    &  $13.678$     \\
UCAC4 592-055831    &$16^{h}33^{m}24^{s}.531$   & $+28^\circ15'54''.818$    &  $14.637$     \\
\hline\hline
\end{tabular}
\end{center}
\end{table}

V1197 Her was also monitored by the 1 meter telescope (YNOs 1m) and the 60 centimeter telescope (YNOs 60cm) at Yunnan Observatories, Chinese Academy of Sciences. A total of two primary (p) minima and four secondary (s) minima were determined by the YNOs 1m and YNOs 60cm. The mid-eclipse times are also listed in Table \ref{New_minimum}.

\begin{table}[!ht]\small
\begin{center}
\caption{New CCD times of mid-eclipse times.}\label{New_minimum}
\begin{tabular}{ccccccc}\hline\hline
    JD (Hel.)     &  Error (days)  &  p/s &           Filter        &       Date        &   Telescope\\\hline\hline
  2457504.3096    & $\pm0.0001$    &   s  &   $B$ $V$ $R_C$ $I_C$   &  25 April 2016    &    TNO 2.4m   \\
  2457505.2231    & $\pm0.0003$    &   p  &           $R_C$ $I_C$   &  26 April 2016    &    YNOs 60cm   \\
  2457517.1818    & $\pm0.0002$    &   s  &           $R_C$         &   8 May 2016      &    YNOs 60cm   \\
  2457855.3733    & $\pm0.0002$    &   p  &       $V$ $R_C$         &  11 April 2017    &    YNOs 60cm   \\
  2457863.3836    & $\pm0.0002$    &   s  &       $V$ $R_C$         &  19 April 2017    &    YNOs 60cm   \\
  2457878.3626    & $\pm0.0001$    &   s  &       $V$ $R_C$ $I_C$   &   4 May 2016      &    YNOs 1m   \\
  2457893.3256    & $\pm0.0002$    &   s  &   $B$ $V$ $R_C$ $I_C$   &  19 May 2016      &    YNOs 60cm   \\
  2458583.2589    & $\pm0.0001$    &   p  &   $B$ $V$ $R_C$ $I_C$   &   9 April 2019    &    TNO 2.4m   \\
\hline\hline
\end{tabular}
\end{center}
\end{table}

\section{INVESTIGATION OF THE PERIOD VARIATION}

The period of close binary is not always constant, especially for semi-detached binary and contact binary due to possible mass transfer between component stars. Thus, the O - C method is used to analysis the period variations of V1197 Her. All available mid-eclipse times are listed Table \ref{Minimum}.

Column 1  - HJD of the observed mid-eclipse times (HJD - 2400000);

Column 2  - primary (p) or secondary (s) mid-eclipse times;

Column 3  - calculated cycle numbers from the initial epoch;

Column 4  - the $O - C$ values calculated from Equation \ref{phase};

Column 5  - observational errors of mid-eclipse times;

Column 6  - vis, pe and CCD refer to visual, photoelectric and Charge Coupled Device observations;

Column 7  - the references;

\begin{small}
\begin{longtable}{llllllll}
\caption{Mid-eclipse times and $O-C$ values for V752 Cen.}\label{Minimum}
\endfirsthead
\multicolumn{7}{l}{Table \ref{Minimum} continued }\\\hline\hline
  JD(Hel.)      &  p/s    &       Epoch       &       $O - C$      &   Error     & Method     & Ref.   \\
 (2400000+)     &         &                   &       (days)       &   (days)    &            &         \\\hline

\endhead
\hline\hline
\endfoot
\endlastfoot
\hline\hline
  JD(Hel.)      &  p/s    &       Epoch       &       $O - C$      &   Error     & Method     & Ref.   \\
 (2400000+)     &         &                   &       (days)       &   (days)    &            &         \\\hline
51967.548 	    &   s	  &     -25185.5      &      0.0162        &   0.008	 &  vis       &    1     \\
51967.681 	    &   p	  &     -25185        &      0.0179        &   0.003	 &  vis       &    1     \\
51984.618 	    &   s	  &     -25120.5      &      0.0120        &   0.004	 &  vis       &    1     \\
52022.448 	    &   s	  &     -24976.5      &      0.0161        &   0.006	 &  vis       &    1     \\
52022.575 	    &   p	  &     -24976        &      0.0118        &   0.005	 &  vis       &    1     \\
52023.625 	    &   p	  &     -24972        &      0.0111        &   0.005	 &  vis       &    1     \\
52041.490 	    &   p	  &     -24904        &      0.0138        &   0.004	 &  vis       &    1     \\
52048.584 	    &   p	  &     -24877        &      0.0155        &   0.002	 &  vis       &    1     \\
52049.373 	    &   p	  &     -24874        &      0.0164        &   0.004	 &  vis       &    1     \\
52049.498 	    &   s	  &     -24873.5      &      0.0101        &   0.006	 &  vis       &    1     \\
52059.477 	    &   s	  &     -24835.5      &      0.0072        &   0.003	 &  vis       &    1     \\
52072.481 	    &   p	  &     -24786        &      0.0086        &   0.007	 &  vis       &    1     \\
52074.459 	    &   s	  &     -24778.5      &      0.0165        &   0.005	 &  vis       &    1     \\
52075.509 	    &   s	  &     -24774.5      &      0.0158        &   0.005	 &  vis       &    1     \\
52117.407 	    &   p	  &     -24615        &      0.0163        &   0.003	 &  vis       &    1     \\
52296.670 	    &   s	  &     -23932.5      &      0.0002        &   0.007	 &  vis       &    1     \\
52323.610 	    &   p	  &     -23830        &      0.0155        &   0.004	 &  vis       &    1     \\
52347.645 	    &   s	  &     -23738.5      &      0.0153        &   0.003	 &  vis       &    1     \\
52367.482 	    &   p	  &     -23663        &      0.0199        &   0.007	 &  vis       &    1     \\
52382.575 	    &   s	  &     -23605.5      &      0.0088        &   0.005	 &  vis       &    1     \\
52409.397 	    &   s	  &     -23503.5      &      0.0375        &   0.005	 &  vis       &    1     \\
52415.552 	    &   p	  &     -23480        &      0.0195        &   0.005	 &  vis       &    1     \\
52438.530 	    &   s	  &     -23392.5      &      0.0130        &   0.006	 &  vis       &    1     \\
52485.408 	    &   p	  &     -23214        &      0.0026        &   0.007	 &  vis       &    1     \\
52702.658 	    &   p	  &     -22387        &      0.0163        &   0.007     &  vis	      &    2    \\
52708.568 	    &   s	  &     -22364.5      &      0.0160        &   0.004     &  vis	      &    2    \\
52764.515 	    &   s	  &     -22151.5      &      0.0121        &   0.006     &  vis	      &    2    \\
52766.490 	    &   p	  &     -22144        &      0.0170        &   0.004     &  vis	      &    2    \\
52813.512 	    &   p	  &     -21965        &      0.0193        &   0.005     &  vis	      &    2     \\
52850.426 	    &   s	  &     -21824.5      &      0.0268        &   0.003     &  vis	      &    3    \\
52867.360 	    &   p	  &     -21760        &      0.0179        &   0.006     &  vis	      &    3    \\
53035.725 	    &   p	  &     -21119        &      0.0050        &   0.005     &  vis	      &    3    \\
53079.604 	    &   p	  &     -20952        &      0.0165        &   0.004     &  vis	      &    3    \\
53082.630 	    &   s	  &     -20940.5      &      0.0216        &   0.003     &  vis	      &    3    \\
53095.629 	    &   p	  &     -20891        &      0.0180        &   0.002     &  vis	      &    3    \\
53112.569 	    &   s	  &     -20826.5      &      0.0151        &   0.003     &  vis	      &    3    \\
53117.554 	    &   s	  &     -20807.5      &      0.0092        &   0.004     &  vis	      &    3    \\
53137.524 	    &   s	  &     -20731.5      &      0.0155        &   0.002     &  vis	      &    3    \\
53150.532 	    &   p	  &     -20682        &      0.0209        &   0.002     &  vis	      &    3     \\
53233.407 	    &   s	  &     -20366.5      &      0.0203        &   0.003     &  vis	      &    4    \\
53406.643 	    &   p	  &     -19707        &      0.0189        &   0.002     &  vis	      &    4    \\
53411.636 	    &   p	  &     -19688        &      0.0209        &   0.002     &  vis	      &    4    \\
53440.544 	    &   p	  &     -19578        &      0.0341        &   0.004     &  vis	      &    4    \\
53445.530 	    &   p	  &     -19559        &      0.0292        &   0.003     &  vis	      &    4    \\
53447.623 	    &   p	  &     -19551        &      0.0208        &   0.002     &  vis	      &    4    \\
53502.393 	    &   s	  &     -19342.5      &      0.0220        &   0.002     &  vis	      &    4    \\
53557.432 	    &   p	  &     -19133        &      0.0295        &   0.004     &  vis	      &    4    \\
53611.406 	    &   s	  &     -18927.5      &      0.0228        &   0.002     &  vis	      &    4    \\
53846.4998	    &   s	  &     -18032.5      &      0.0180        &   0.0004    &  pe        &    5    \\
54556.9209	    &   p	  &     -15328        &      0.0210        &   0.0001    &  CCD       &    6    \\
55643.8867	    &   p	  &     -11190        &      0.0170        &   0.0001    &  CCD       &    7    \\
57156.7867	    &   s	  &     -5430.5       &      0.0115        &   0.0002    &  CCD       &    8    \\
57504.3096      &   s     &     -4107.5       &      0.0088        &   0.0001    &  CCD       &    9    \\
57505.2231      &   p     &     -4104         &      0.0029        &   0.0003    &  CCD       &    9    \\
57517.1818      &   s     &     -4058.5       &      0.0097        &   0.0002    &  CCD       &    9    \\
57855.3733      &   p     &     -2771         &      0.0007        &   0.0002    &  CCD       &    9    \\
57863.3836      &   s     &     -2740.5       &      -0.0008       &   0.0002    &  CCD       &    9    \\
57878.3626      &   s     &     -2683.5       &      0.0055        &   0.0001    &  CCD       &    9    \\
57893.3256      &   s     &     -2626.5       &      -0.0043       &   0.0002    &  CCD       &    9    \\
58583.2589      &   p     &     0             &      0             &   0.0001    &  CCD       &    9    \\\hline
\end{longtable}
\textbf
{\footnotesize References:} \footnotesize (1) BBSAG Bulletins\footnote{https://www.astronomie.info/bbsag/bulletins.html}; (2) \citet{2003IBVS.5438....1D}; (3) \citet{2004IBVS.5543....1D}; (4) \citet{2005OEJV....3....1L}; (5) \citet{2006IBVS.5713....1D}; (6) \citet{2009IBVS.5875....1N}; (7) \citet{2012IBVS.6018....1N}; (8) \citet{2016IBVS.6164....1N};  (9) The present work.
\end{small}

\begin{figure}[!ht]
\begin{center}
\includegraphics[width=12cm]{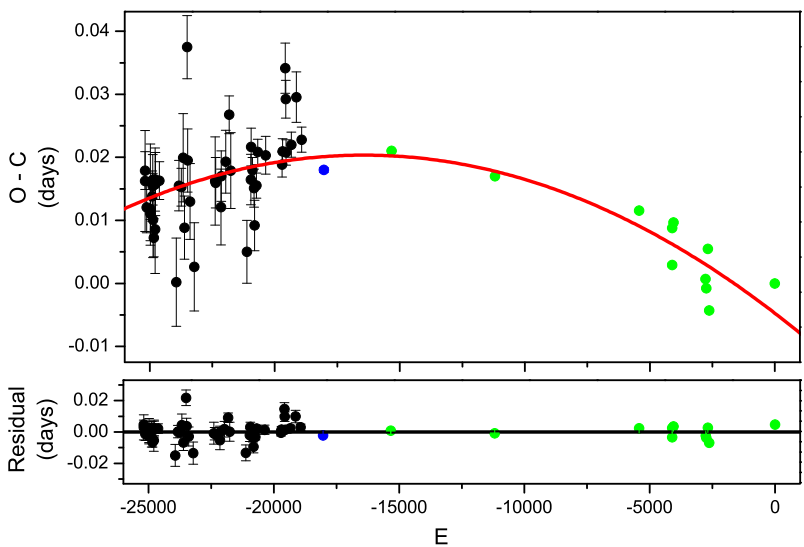}
\caption{The black, blue and green colors refer to mid-eclipse times obtained by visual, photoelectric and Charge Coupled Device observations, respectively. The red line represents theoretical O - C curve based on Equation \ref{New_ephemeris}. }\label{O-C}
\end{center}
\end{figure}

As displayed in Fig. \ref{O-C}, the period of V1197 Her is not constant. Therefore, parabola fit is performed on the data and a new ephemeris is determined:
\begin{equation}\label{New_ephemeris}
\begin{array}{lll}
Min. I = 2458583.25417(\pm0.00006)+0.26267695(\pm0.00000001)\times{E}
         \\-9.28(\pm0.06)\times{10^{-11}}\times{E^{2}}
\end{array}
\end{equation}

The newly determined ephemeris reveals that the period of V1197 Her is decreasing continuously at a rate of $dP/dt=-2.58\times{10^{-7}}day\cdot year^{-1}$.

\section{MODELING THE LIGHT CURVES}

\begin{figure}[!ht]
\begin{center}
\includegraphics[width=12cm]{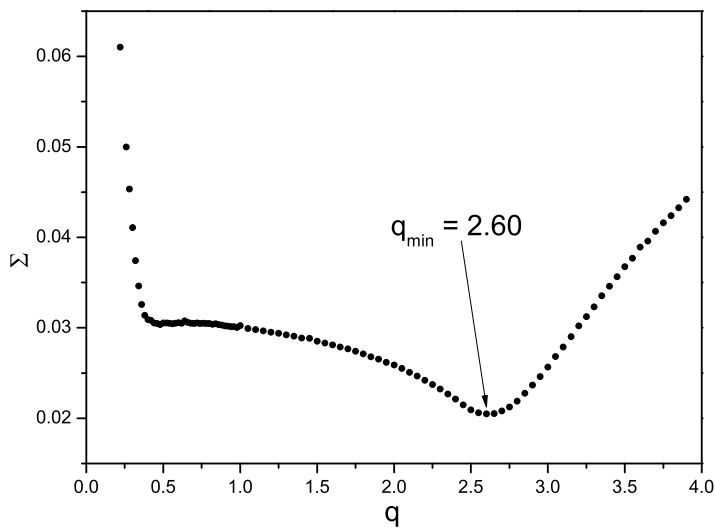}
\caption{The q-search diagram.}\label{q-search}
\end{center}
\end{figure}

\begin{figure}[!ht]
\begin{center}
\includegraphics[width=12cm]{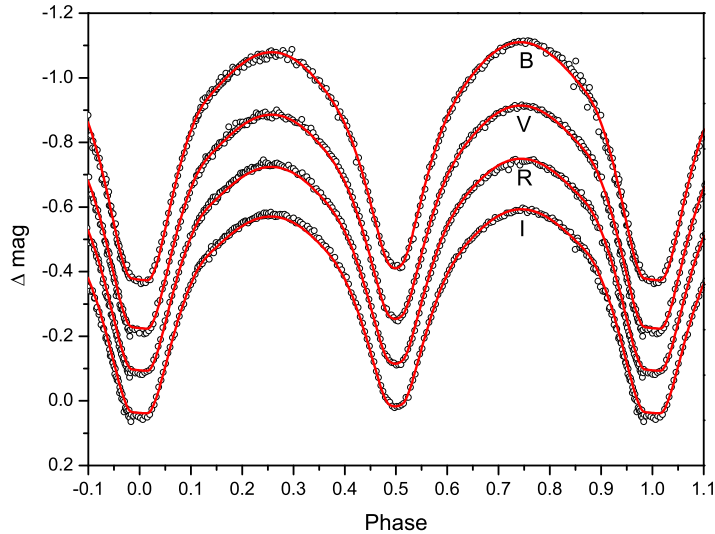}
\caption{The open circles are observational light curves obtained using the TNO 2.4m. The red lines are theoretical light curves calculated with the Wilson-Devinney program.}\label{LC_Cal}
\end{center}
\end{figure}

\begin{figure}[!ht]
\begin{center}
\includegraphics[width=13cm]{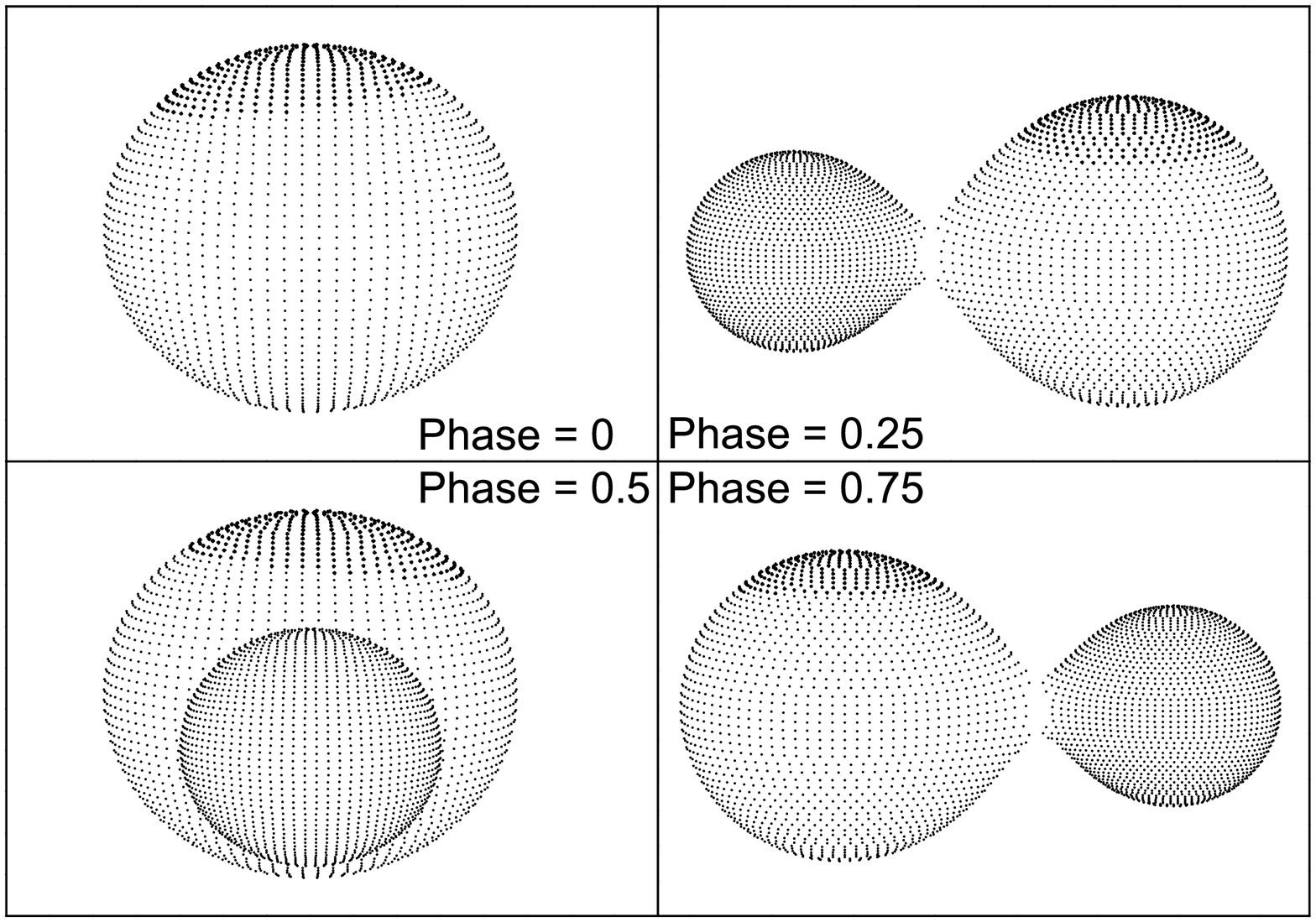}
\caption{The geometrical structure at Phase = 0, Phase = 0.25, Phase = 0.5 and Phase = 0.75.}\label{CF}
\end{center}
\end{figure}

V1197 Her is a newly reported eclipsing binary. Analyses of its light curves have been neglected and it is the first time that complete multi-color light curves are obtained. To get the physical parameters of V1197 Her, the Wilson-Devinney (W-D) program \citep{1971ApJ...166..605W,2012AJ....144...73W} is applied to model the light curves displayed in Fig. \ref{LC_Obs}. V1197 Her shows EW type light curves. Thereforce, Mode 3 for overcontact binaries is selected. Mean surface temperature of the secondary star (star 2) is fixed as $T_2 = 4973K$ basing on the Gaia Data Release 2 (Gaia DR2) \citep{2016A&A...595A...1G,2018A&A...616A...1G}, and the star eclipsed at primary minimum is assumed as primary star. The free parameters include mass ratio ($q = M_2/M_1$), orbital inclination ($i$), modified dimensionless surface potential of star 1 ($\Omega_{1}$), mean surface temperature of star 1 ($T_{1}$), bandpass luminosities of star 1 ($L_{1}$). As the light curve is asymmetric, the latitude, longitude, angular radius and dimensionless temperature factor for spot are also adjustable. At first, we run the W-D program with a fixed q value (q-search method) to find a proper mass ratio for the binary system. The results suggest that the best fit mass ratio is $q = 2.60$ as displayed in Fig. \ref{q-search}. Then, we set mass ratio as a free parameter together with other parameters and give the initial value as $q = 2.60$ to find a convergent solution. The final solutions for all elements are listed in Table \ref{WD_results} and the theoretical light curves are plotted in Fig. \ref{LC_Cal}. The geometrical structure of V1197 Her at phase 0, 0.25, 0.50 and 0.75 are shown in Fig. \ref{CF}. The location of spot we added in star 2 is easily to be seen in Fig. \ref{CF}. And also, we have tried to set third light ($l_3$) as a free parameter when running the W-D program since tertiary components are commonly reported in contact binaries \citep{2016PASP..128d4201Y,2019MNRAS.485.4588L}. However, no convergent solution is acquired in that case.

\begin{table}[!h]
\begin{center}
\caption{Photometric solutions for V1197 Her}\label{WD_results}
\small
\begin{tabular}{lllllllll}
\hline\hline
Parameters                            &   Values                      \\\hline
$T_{1}(K)   $                         &  5113($\pm5$)                  \\
q ($M_2/M_1$ )                        &  2.61($\pm0.04$)                 \\
$i(^{\circ})$                         &  82.7($\pm0.3$)                    \\
$T_{2}(K)$                            &  4973(fixed)                       \\
$\Delta T(K)$                         &  140($\pm5$)                         \\
$T_{2}/T_{1}$                         &  0.973($\pm0.001$)                 \\
$L_{1}/(L_{1}+L_{2}$) ($B$)           &  0.3408($\pm0.0008$)                \\
$L_{1}/(L_{1}+L_{2}$) ($V$)           &  0.3316($\pm0.0008$)                 \\
$L_{1}/(L_{1}+L_{2}$) ($R_c$)         &  0.3247($\pm0.0012$)                \\
$L_{1}/(L_{1}+L_{2}$) ($I_c$)         &  0.3203($\pm0.0014$)                  \\
$r_{1}(pole)$                         &  0.287($\pm0.001$)                  \\
$r_{1}(side)$                         &  0.300($\pm0.002$)                   \\
$r_{1}(back)$                         &  0.338($\pm0.002$)                    \\
$r_{2}(pole)$                         &  0.443($\pm0.005$)                    \\
$r_{2}(side)$                         &  0.475($\pm0.007$)                    \\
$r_{2}(back)$                         &  0.504($\pm0.010$)                    \\
$\Omega_{1}=\Omega_{2}$               &  6.00($\pm0.06$)                     \\
$f$                                   &  $15.7\,\%$($\pm$9.2\,\%$$)      \\
$\theta(^{\circ})$                    &  6.5($\pm1.4$)                    \\
$\psi(^{\circ})$                      &  106.2($\pm3.0$)                    \\
$r$(rad)                              &  0.75($\pm0.07$)                    \\
$T_f$                                 &  0.82($\pm0.07$)                      \\
$\Sigma{\omega(O-C)^2}$               &  0.0179                                 \\
\hline
\hline
\end{tabular}
\end{center}
\end{table}

\section{DISCUSSION AND CONCLUSION}

\begin{figure}[!ht]
\begin{center}
\includegraphics[width=13cm]{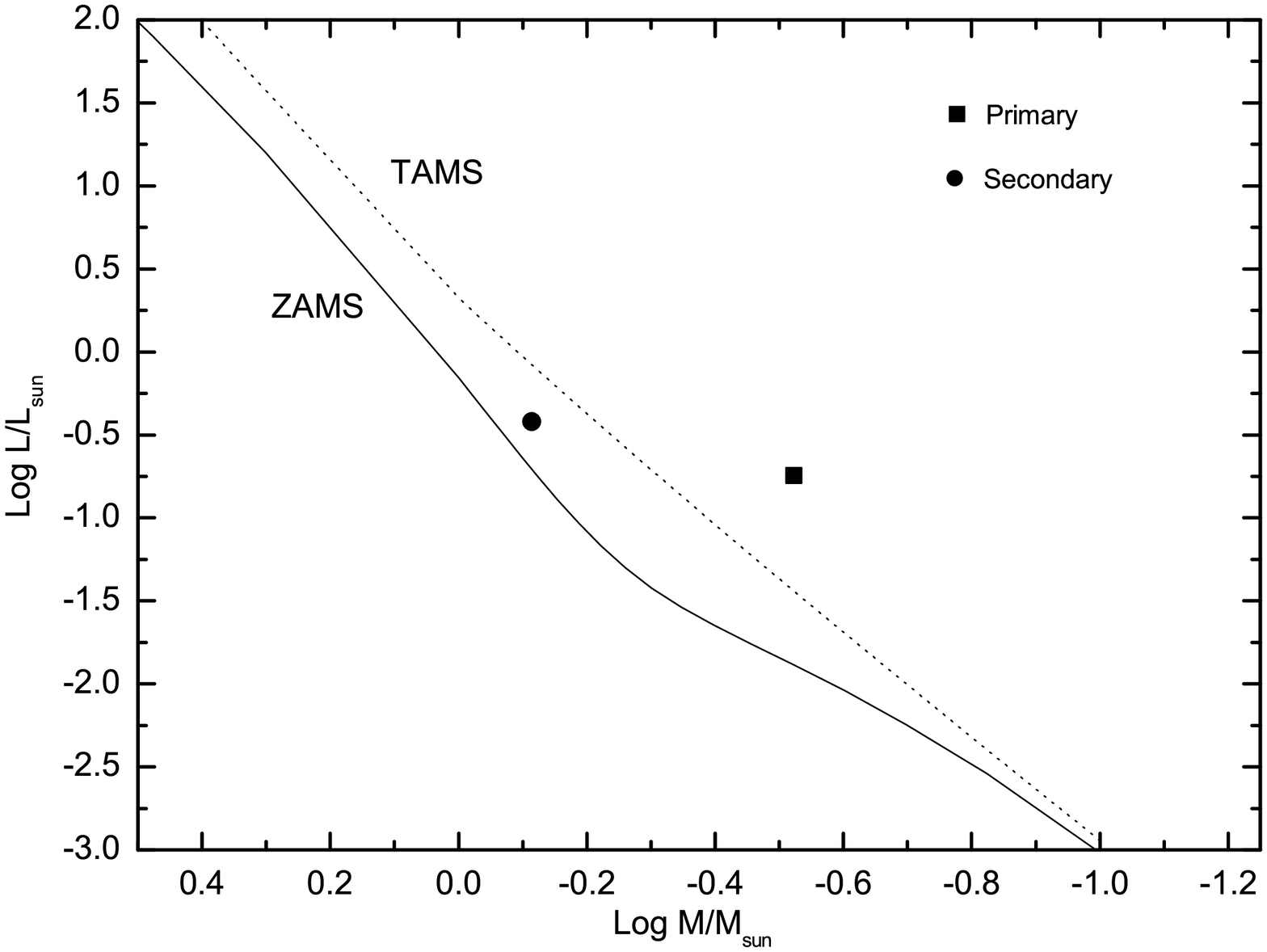}
\caption{The H - R diagram.}\label{H-R}
\end{center}
\end{figure}

The photometric solutions for V1197 Her are determined for the first time basing on the multi-color light curves obtained with the TNO 2.4m telescope. According to the results listed in Table \ref{WD_results}, V1197 Her is a W-subtype shallow contact binary with its mass ratio to be $q = 2.61(4)$ and fill-out factor to be $f = 15.7(9.2)\,\%$. The primary star (star 1) is $140(5)K$ hotter than the more massive secondary star (star 2). V1197 Her has a very high orbital inclination ($i = 82.7^{\circ} $) and the geometrical structure at Phase = 0 (Fig. \ref{CF}) shows that the primary star is totally eclipsed by the secondary star, which means that V1197 Her is a completely eclipsing binary system and the determined parameters in Table \ref{WD_results} are very reliable \citep{2005Ap&SS.296..221T,2018PASP..130g4201L,2019RAA....19...56H}. Basing on the mean surface temperature of star 2 ($T_{2} = 4973K$), its mass is estimated to be $ M_2 = 0.77(2) M_\odot$ \citep{Cox2000}. Then, the masses, radii and luminosities of the two component stars in V1197 Her are calculated, which are listed in Table \ref{absolute}. And the orbital semi-major axis of this binary system is calculated to be $a = 1.76(2)R_\odot$. The evolutionary status of the primary star and secondary star are plotted in the Hertzsprung-Russell (H-R) diagram (Fig. \ref{H-R}), which implies that the secondary star is still a main sequence star while the primary star has evolved away from the main sequence stage. The period variations of V1197 Her is revealed to be decreasing continuously at a rate of $dP/dt=-2.58\times{10^{-7}}day\cdot year^{-1}$, which may due to the mass transfer from the more massive star (star 2) to the less massive one (star 1) \citep{2016AJ....152..120H,2019AJ....157..207L}. The mass transfer rate is estimated to be $\frac{dM_{2}}{dt}=- 1.61\times{10^{-7}}M_\odot/year$.

\begin{table}[!h]
\caption{Absolute parameters of components in V1197 Her}\label{absolute}
\begin{center}
\small
\begin{tabular}{lllllllll}
\hline
Parameters                        &Primary (star1)                 & Secondary (star 2)         \\
\hline
$M$                               & $0.30(\pm0.01)M_\odot$         & $0.77(\pm0.02)M_\odot$         \\
$R$                               & $0.54(\pm0.01)R_\odot$         & $0.83(\pm0.01)R_\odot$         \\
$L$                               & $0.18(\pm0.01)L_\odot$         & $0.38(\pm0.01)L_\odot$         \\
\hline
\end{tabular}
\end{center}
\end{table}

The negative O'Connell effect is observed to appear on the light curves of V1197 Her, in which the light maximum at phase 0.75 (Max II) is brighter than the light maximum at phase 0.25 (Max I). The  magnitude difference between the two light maxima (Max I - Max II) are 0.031 mag in $B$ band, 0.027 mag in $V$ band, 0.022 mag in $R_c$ band and 0.019 mag in $I_c$ band. It is suggested that magnetic activity of the component stars mainly account for the O'Connell effect on solar-type or late-type contact binaries \citep{2014ApJS..212....4Q,2016PASJ...68..102X,2017PASJ...69...79L}. Some interesting characteristics of the O'Connell effect are reported recently. The O'Connell effect appeared on UCAC4436-062932 in a very short timescale \citep{2016NewA...47....3Z} and the O'Connell effect even changed from a negative one to a positive one or otherwise \citep{2018PASJ...70...87Z,2019PASJ...71...39Z}. Compared with the O'Connell effect on early-type contact binaries, the magnitude differences (Max I - Max II) of the O'Connell effect on late-type contact binaries is much smaller, usually less than 0.04 mag \citep{2011AN....332..607P}.

The formation and evolution of contact binaries is still an open issue. They may evolve from initially detached binary systems due to angular momentum loss \citep{1994ASPC...56..228B,2007ApJ...662..596L}. Contact binaries are classified into A-subtype or W-subtype systems \citet{1970VA.....12..217B}. \citet{2008MNRAS.387...97L} thought that there was no evolutionary difference between A-subtype and W-subtype systems, and A- and W-subtype systems might be just in different stages of thermal relaxation oscillation(TRO). However, \citet{2013MNRAS.430.2029Y} claimed that initial masses of the secondary star in a binary system is the key factor to affect its evolutionary status. Binary systems with initial masses of the secondary stars higher than $1.8(\pm0.1)M_\odot$
will evolve into A-subtype contact binaries while binaries with initial masses lower than this evolve into W-subtype systems. Therefore, more and more observations and analyses on both A- and W-subtype contact binary systems are needed to understand the formation mechanism of W UMa-type contact binaries.

\begin{acknowledgements}
We appreciate the valuable comments and suggestions from the anonymous referee. This research was supported by the Chinese Natural Science Foundation (Grant No. 11703080 and 11703082) and the Yunnan Natural Science Foundation (No. 2018FB006). It was part of the research activities at the National Astronomical Research Institute of Thailand (Public Organization). This work has made use of data from the European Space Agency (ESA) mission {\it Gaia} ({https://www.cosmos.esa.int/gaia}), processed by the {\it Gaia} Data Processing and Analysis Consortium (DPAC, {https://www.cosmos.esa.int/web/gaia/dpac/consortium}). Funding for the DPAC has been provided by national institutions, in particular the institutions participating in the {\it Gaia} Multilateral Agreement.
\end{acknowledgements}

\label{lastpage}

\end{document}